\documentclass[sigconf]{acmart}
\renewcommand\footnotetextcopyrightpermission[1]{} 
\usepackage{svg}
\usepackage{graphicx}
\usepackage{textcomp}
\usepackage{xcolor}
\usepackage{threeparttable,tabularx}
\usepackage[normalem]{ulem}
\useunder{\uline}{\ul}{}

\usepackage{cases}

\usepackage{verbatim}
\usepackage{amsmath,amssymb,amsfonts} 
\usepackage{subfigure}
\usepackage{threeparttable}

\usepackage{booktabs}
\usepackage{multirow}
\usepackage{algorithm}  
\usepackage{algorithmicx}  
\usepackage{algpseudocode}
\AtBeginDocument{%
  \providecommand\BibTeX{{%
    \normalfont B\kern-0.5em{\scshape i\kern-0.25em b}\kern-0.8em\TeX}}}

\acmSubmissionID{100}

\begin{document}
\begin{sloppypar}

\title{FHPM: Fine-grained Huge Page Management For Virtualization}

\author{Chuandong Li}
\affiliation{%
    \institution{Peking University}
    \country{}
}
\author{Sai Sha}
\affiliation{%
    \institution{Peking University}
    \country{}
}
\author{Yangqing Zeng}
\affiliation{%
    \institution{Peking University}
    \country{}
}
\author{Xiran Yang}
\affiliation{%
    \institution{Peking University}
    \country{}
}
\author{Yingwei Luo}
\affiliation{%
    \institution{Peking University}
    \country{}
}
\author{Xiaolin Wang}
\affiliation{%
    \institution{Peking University}
    \country{}
}
\author{Zhenlin Wang}
\affiliation{%
    \institution{Michigan Technological University}
    \country{}
}


\begin{abstract}

As more data-intensive tasks with large footprints are deployed in virtual machines (VMs), huge pages are widely used to eliminate the increasing address translation overhead. However, once the huge page mapping is established, all the base page regions in the huge page share a single extended page table (EPT) entry, so that the hypervisor loses awareness of accesses to base page regions. None of the state-of-the-art solutions can obtain access information at base page granularity for huge pages. We observe that this can lead to incorrect decisions by the hypervisor, such as incorrect data placement in a tiered memory system and unshared base page regions when sharing pages.

This paper proposes {\it FHPM}, a fine-grained huge page management for virtualization without hardware and guest OS modification. {\it FHPM} can identify access information at base page granularity, and dynamically promote and demote pages. A key insight of {\it FHPM} is to redirect the EPT huge page directory entries (PDEs) to new companion pages so that the MMU can track access information within huge pages. Then, {\it FHPM} can promote and demote pages according to the current hot page pressure to balance address translation overhead and memory usage. At the same time, {\it FHPM} proposes a VM-friendly page splitting and collapsing mechanism to avoid extra VM-exits. In combination, {\it FHPM} minimizes the monitoring and management overhead and ensures that the hypervisor gets fine-grained VM memory accesses to make the proper decision. Our evaluation shows that {\it FHPM} can accurately monitor the accesses of base page regions within a huge page, while incurring only a performance penalty of less than 4\%. Moreover, the VM-friendly page splitting and collapsing  is 60\% faster than that of the native Linux interface. We apply {\it FHPM} to improve tiered memory management ({\it FHPM-TMM}) and to promote page sharing ({\it FHPM-Share}). {\it FHPM-TMM} achieves a performance improvement of up to 33\% and 61\% over the pure huge page and base page management. {\it FHPM-Share} can save 41\% more memory than Ingens, a state-of-the-art page sharing solution,  with comparable performance.  
\end{abstract}



\keywords{Virtual Machine, Huge Page Management, Extended Page Table, Tiered Memory Management, Page Sharing}



\maketitle
\pagestyle{plain} 

\section{Introduction}

As the foundation of cloud computing, virtualization is widely used by cloud service providers to manage cloud resources to ensure resource sharing, security, and isolation~\cite{bhardwaj2010cloud}. Cloud services are typically leased to cloud users in the form of virtual machines (VMs)~\cite{MargaritovUSG21}. In-memory databases, data analysis frameworks, and graph processing programs are commonly run in VMs, which have increasingly large memory requirements, with terabyte-scale applications emerging in recent years~\cite{scaleoutworkloads,themachine}. For those applications, the two-dimensional address translation due to virtualization remains a performance bottleneck. For a memory access, guest virtual addresses (GVAs) must be translated to guest physical addresses (GPAs), and then to host physical addresses (HPAs). The most widely used hardware-assisted memory virtualization is Intel Extended Page Table (EPT)~\cite{intelept} and AMD Nested Page Table (NPT)~\cite{amdnpt} (We use EPT in this paper). The host maintains the page table mapping from GPAs to HPAs in the EPT, and the hardware performs a long latency traversal of the guest page table and the EPT when a translation look-aside buffer (TLB) miss occurs. The growing memory footprint of applications also puts significant pressure on the TLB capacity. A significant number of TLB misses can lead to high latency page walks, thereby causing severe degradation of application performance in cloud.


Huge pages are often chosen to speed up address translation due to its several advantages. First, a single page fault is generated for every huge page virtual region, thereby reducing the frequency of enter/exit kernel operations. Second, the elimination of the last level structure of the page table, especially when using EPT for virtualization, reduces the number of memory accesses in address translation from 24 to 15. Finally, a single TLB entry maps a much larger amount of virtual memory, thereby reducing the number of TLB misses~\cite{hugerst}.

However, aggressive use of huge pages is not a panacea. Huge pages can result in increased page fault latency, memory bloat and memory fragmentation. To address these issues, several mature solutions have been proposed, such as FreeBSD~\cite{freebsd}, Ingens~\cite{ingens}, Hawkeye~\cite{hawkeye}, Quicksilver~\cite{quicksilver}. We noticed that those issues arises in situations with frequent memory allocation and deallocation. However, in cloud computing environments, the secondary mapping of VMs is stable, and the host OS does not actively free the VM pages, thereby avoiding the memory bloat or fragmentation of the host memory. 

In virtualization, the use of huge pages presents a significant challenge due to the loss of access information on base page granularity. All base page regions within a huge page share the accessed/dirty bits (A/D bits) in the PDE, which results in the hypervisor being unable to control and monitor the base pages individually. Scanning A/D bits in the EPT is commonly employed in VM memory management~\cite{hirofuchi2016raminate, idlepage, vmware15vee, Waldspurger02} due to its low overhead. The use of huge pages, with coarse granularity, imposes limitations on the flexibility of memory management, leading to suboptimal performance and resource utilization. For instance, the use of huge pages can negatively impact optimal placements of data in tiered memory management systems or NUMA memory management systems~\cite{GaudLDFFQ14}. Additionally, the increased page size can reduce the probability of page sharing opportunities by $4-10 \times$~\cite{micro15canyou}. VMware and Ingens demote the huge pages that are infrequently  accessed by checking their A/D bits\cite{vmware15vee, ingens}. But the coarse-grained access information still impedes the optimal decision-making.



In this paper, we provide a comprehensive analysis of the performance degradation and low memory utilization that can result from tracking huge pages. Our study reveals that the use of huge pages causes the hypervisor to lose awareness of the access information of base page granularity, leading to suboptimal decisions in different memory management scenarios based on page tracking. We attribute this issue to the presence of {\it unbalanced huge pages}, where accesses to a huge page are concentrated on only a few base page regions within the huge page. This can mislead the hypervisor into treating the entire huge page as hot, which we refer to as {\it hot bloat}. We propose Fine-grained Huge Page Management (FHPM) to monitor the memory access at base page granularity with low overhead and dynamically promote and demote pages to balance address translation overhead and memory usage. FHPM maintains a companion page for each PDE of huge pages and replaces it with an entry referring to the companion page when monitoring. Once the monitoring is complete, it gracefully returns to the original state by recovering the initial PDE. We then dynamically promote and demote pages based on the access pattern at the base page granularity to enable proper decision-making in different scenarios. In addition, page splitting and collapsing results in additional VM-exits and thus higher performance overhead compared to the native environment. Therefore, we implement FHPM with VM-friendly page splitting and collapsing, by refilling the EPT entry after page remapping. We implemented FHPM in a tiered memory management system and a page sharing system to show the improvement of performance and memory utilization.

In summary, our work makes the following contributions:
\begin{itemize} 
\item We identify and analyze the phenomenon of {\it hot bloat}, which causes the hypervisor to make suboptimal decisions and leads to degradation of performance and memory utilization.
\item We propose a novel companion page design to monitor memory access pattern at base page granularity with low overhead and without guest OS and hardware modifications.
    \item We demote and promote pages dynamically according to the hot page pressure of the VM in various scenarios to balance the address translation overhead and memory utilization.
	\item We propose a VM-friendly page splitting and collapsing mechanism to minimize the extra VM-exits when demoting and promoting pages.
	\item We implement and evaluate FHPM in two cases, tiered memory management and memory sharing, to show its superiority and compare it with the state-of-the-art designs.
	
\end{itemize} 
\section{Background and Related Works}
In this section, we provide an overview of the current huge page management systems, and a brief introduction to the tiered memory management system and page-sharing system, which FHPM will later be applied to. 

\subsection{Huge Page Management}
The huge page is often used to mitigate the two-dimensional address translation overhead of VMs. We observe that using huge pages in virtualization has a significant performance improvement across various programs with diverse memory access patterns on distinct memory mediums. For instance, our experiments demonstrate that enabling huge pages in \texttt{Redis} (we test all the benchmarks in §\ref{sec:ex-setup}, the results of other benchmarks exhibit similar trends) leads to performance gains of 11.94\%, 6.28\%, and 5.54\% for local DRAM, remote DRAM, and NVM, respectively. To further evaluate the impact of using huge pages, we employ hardware performance monitoring unit to measure the number of TLB misses and page walk latency. Our findings demonstrate that the utilization of huge pages can reduce the number of TLB misses by an order of magnitude and decrease page walk latency by up to 40\%.

Numerous studies have highlighted several issues associated with the use of huge pages, such as high first-touch page fault latency, memory bloat, and memory fragmentation. Linux THP~\cite{thp} allocates, prepares and maps huge pages synchronously on the first page fault and deallocates them when partially freed. FreeBSD~\cite{freebsd} and Quicksilver~\cite{quicksilver} decouple physical huge page allocation from preparation using a reservation-based allocator. They create huge page mapping after all or some base page mappings are established and defer huge page deallocation as long as possible. Additionally, Quicksilver provides a reservation page recycle mechanism to prevent bloat and fragmentation. Ingens~\cite{ingens} allocates, prepares, and maps a base page on the first page fault and creates a huge page mapping only when a sufficient number of base pages are established. When under memory pressure, Ingens demotes cold huge pages by scanning A/D bits to share base pages. HawkEye~\cite{hawkeye} uses asynchronous preparation and creates huge page mapping directly. When under memory pressure, HawkEye demotes huge pages to recycle zero base pages. 
Also, many hardware solutions have been proposed to address the issue of allocating huge pages in non-contiguous physical memory, including Perforated Page~\cite{0001CKKBH20}, GTSM~\cite{DuZCMM15}, and GLUE~\cite{PhamVLB15}. However, these solutions solely focus on addressing the issues of frequent memory allocation and deallocation, without providing a mechanism for acquiring internal access information of huge pages.

\subsection{Tiered Memory Management}
VMs are increasingly demanding in terms of memory requirements, but scaling memory for cloud service providers is limited by the capacity and power consumption of DRAM. Typically, the tiered memory system is partitioned into fast memory, such as DRAM, with higher bandwidth and lower access latency but limited capacity, and slow memory, such as non-volatile memory (NVM)\cite{2018Big,lee2009architecting, apalkov2013spin} and Compute Express Link (CXL) memory\cite{cxl}, with higher capacity but higher latency and lower bandwidth.

A key performance factor of a tiered memory management system is to place frequently accessed data (i.e., ``hot'' data) in fast memory and infrequently accessed data (i.e., ``cold'' data) in slow memory to ensure an acceptable performance loss while expanding memory capacity. To achieve this, page tracking, classification, and migration are employed. Various methods are used for page tracking, including processor event-based sampling (PEBS), used by HeMem~\cite{raybuck2021hemem}, write protection, used by Thermostat~\cite{thermostat} and Autotiering~\cite{autotiering}, and page table scan, employed by DAMON~\cite{damon}, Raminate~\cite{RAMinate}, HMM-v~\cite{sha2022hmmv}, and HeteroOS~\cite{heteroos}. For hot page classification, HeMem and Thermostat classify pages with a fixed frequency, while Raminate, Nimble, and HeteroOs use the LRU algorithm. HMM-v employs a bucket sort algorithm to select top accessed pages as hot pages. And for page migration, Nimble support huge page migration in parallel, HeMem use write-protection to recopy dirty pages, and HMM-v use page modification logging to recopy dirty pages in virtualization. Among these solutions, HMM-v is the state-of-the-art tiered memory management system customized for virtualization and serves as our comparison baseline in this study. Accurate page tracking is essential for optimal data placement in all these approaches. However, PEBS lacks virtualization support, and write protection in virtualization is expensive, leading to the adoption of page table scanning, wherein access/dirty (A/D) bits in the EPT are periodically cleared and recorded, in tiered memory management for virtualization.
In §\ref{sec:motivation}, we demonstrate that current tiered memory management based on page table tracking can cause hot data to be placed in slow memory, leading to performance degradation when huge pages are enabled.

\subsection{Page Sharing}
\label{sec:pageshare}
Page sharing techniques are often employed to reduce memory consumption and improve physical memory utilization. One such technique is kernel same-page merging (KSM)~\cite{ksmrst}, which transparently shares pages at the base page granularity. KSM uses a ``share first'' strategy and splits huge pages once a base page region can be shared within them. Another method is huge page sharing~\cite{Waldspurger02}, which only shares huge pages with the same content without splitting them. However, this approach is not very effective in practice. VMware~\cite{vmware15vee} and Ingens~\cite{ingens} use a trade-off approach, periodically checking the access/dirty (A/D) bits of pages, keeping hot huge pages to avoid address translation overhead, and splitting cold huge pages to promote base-page sharing for more memory space. However, these strategies rely on huge page table scan, which may not accurately track memory accesses. In §\ref{sec:motivation}, we demonstrate that the current page sharing system cannot achieve an optimal balance between memory savings and address translation overhead when huge pages are enabled. 
\section{The challenges of huge page management under virtualization}
\label{sec:motivation}
Under virtualization, the hypervisor allocates physical memory to each VM upon first touched, and does not automatically reclaim memory thereafter. The second-level mapping from GPAs to HPAs remains stable at the hypervisor level. The host OS considers the memory occupied by the VM to be in use, even if the application inside the VM has been deallocated. Therefore, memory bloat and fragmentation are not caused by the host OS. From the VM perspective, the aforementioned solutions can still be employed within the guest OS to mitigate memory-related issues and avoid additional memory requests to cloud providers, which can incur additional costs. The cloud service provider is not concerned with memory bloat and fragmentation within the VM because those memory requests are equally charged. They are more interested in identifying under-utilized memory resources to improve overall resource utilization and reduce costs. However, using huge pages causes the hypervisor to lose awareness on base page granularity. In this section, we analyze the challenges brought by huge pages in virtualization.

\subsection{Hot Bloat}
\label{sec:hotpage}
\begin{figure}[htbp]
\begin{center}
\includegraphics[width=0.48\textwidth]{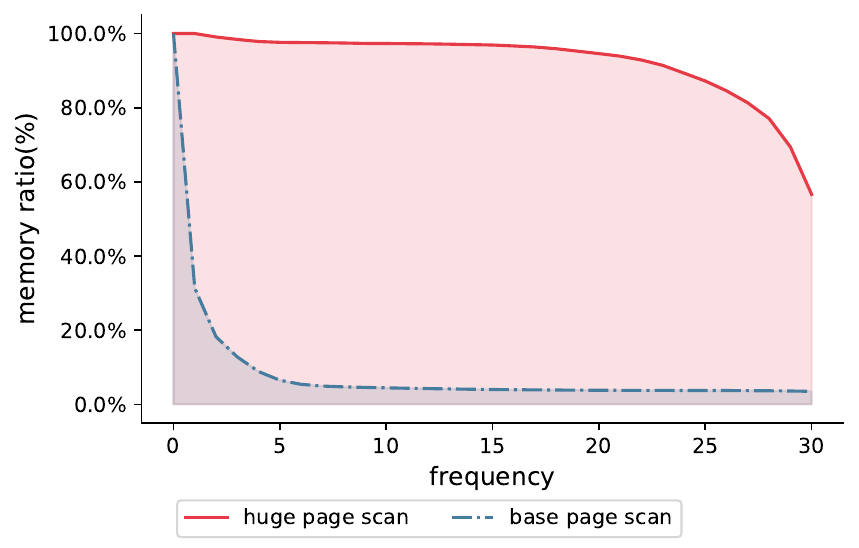}
\caption{The CCDF graph of memory access frequency of base page scan and huge page scan. The x-axis represents access frequency, and the y-axis represents the proportion of memory with access frequency greater than x.
}   
\label{fig:redis-hugebase}
\end{center}
\vspace{-0.5cm}
\end{figure}

When using base pages, the hypervisor can periodically clear the A/D bits in the EPT entry to obtain the memory access information at base page granularity in the VM. However, when using huge pages, all base pages within a huge page share the A/D bits in the PDE. Consequently, the hypervisor cannot determine which base page regions within a huge page have been accessed when the A/D bits are set. To demonstrate this issue, we conducted experiments on an 8-core 32 GB VM running benchmarks in §\ref{sec:ex-setup}. We monitored the VM memory accesses by scanning the A/D bits in the EPT with the Linux THP enabled and disabled, respectively, and with one scan per second for 30 rounds of scanning. The results of \texttt{Redis} are presented here, and similar trends are observed in the other benchmarks. The resulting complementary cumulative distribution function (CCDF) plot of memory access frequency, shown in Figure~\ref{fig:redis-hugebase}, indicates that the memory access frequency is completely opposite when using huge pages and base pages, respectively.

We define an {\it unbalanced huge page} as a huge page in which only a few base page regions are frequently accessed. In contrast, we refer to a huge page where most parts are accessed as a {\it balanced huge page}. Due to the limitations of the conventional page-based memory access tracking methods, such as page table scanning and page fault tracking, accessing only a few base page regions of an {\it unbalanced huge page} is considered as accessing the entire huge page and therefore the OS cannot differentiate between an {\it unbalanced huge page} and a {\it balanced huge page}. {\it Unbalanced huge pages} result in the difficulty of accurately distinguishing between hot and cold data, leading to a false indication of global heat for the entire VM/application, a phenomenon termed {\it hot bloat}. To quantify the distribution of accesses to base page regions in a huge page, we introduce the Page Skew Ratio (PSR), defined as $$ PSR = 1 - \frac{N_s}{N_h},$$ where $N_s$ is the number of base page regions actually accessed in a huge page and $N_h$ is the total number of base pages contained in a huge page (512 in the case of Linux). A large PSR on a huge page indicates that accesses to that page are skewed towards accessing a few base pages within it. Table~\ref{tab:PSR} shows the PSR distribution of Redis, with a significant proportion of high PSR pages, which explains why the access characteristics of huge pages differ from those of base pages. The problem of memory bloat in a VM is a special case of hot bloat, where a huge page is produced with only a few base page regions in use and the remainder completely untouched. 

{\it hot bloat} can mislead the hypervisor into making inappropriate decisions. Next, we use microbenchmarks to briefly illustrate the sub-optimal results in tiered memory management and page sharing.

\begin{table}[htbp]
\centering
\tabcolsep=0.1cm
  \caption{Page Skew Ratio distribution of Redis.}
  \vspace{-0.3cm}
\begin{tabular}{cc|cc}
\hline
PSR & huge page size & PSR & huge page size \\ \hline
{[}0,0.1)       & 1124             & {[}0.5,0.6)     & 10               \\
{[}0.1-0.2)     & 4                & {[}0.6,0.7)     & 14               \\
{[}0.2,0.3)     & 0                & {[}0.7,0.8)     & 938              \\
{[}0.3,0.4)     & 0                & {[}0.8,0.9)     & 1754             \\
{[}0.4,0.5)     & 4                & {[}0.9,1.0)     & 20770            \\ \hline
\end{tabular}
\label{tab:PSR}
\end{table}

\subsection{Performance Loss in Tiered Memory Management}
\begin{figure}[ht]
\begin{center}
\includegraphics[width=0.48\textwidth]{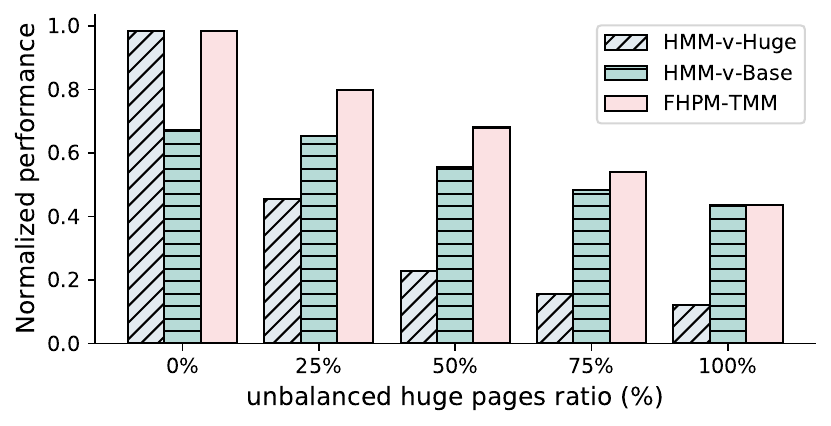}
\caption{The normalized performance of different proportions of unbalanced huge pages for HMM-v-Huge, HMM-v-Base, and FHPM-TMM. The baseline is enabling huge pages with sufficient fast memory. }
\label{fig:microvtmm}
\end{center}
\vspace{-0.5cm}
\end{figure}

To demonstrate the impact of using huge pages on the tiered memory management system, we utilize the state-of-the-art approach, HMM-v, under pure huge pages and base pages management (HMM-v-Huge and HMM-v-Base). We configure the VM with 8 cores, 8GB of DRAM as fast memory, and 120GB of NVM as slow memory. The micro-benchmark is a 40GB random access program with a 1:1 read and write ratio, focusing on accessing 4GB memory. We vary the proportion of unbalanced huge pages and fix the PSR of these pages at 0.9 while setting the PSR of the remaining balanced huge pages at 0 (i.e., every base page region is accessed). As shown in Figure~\ref{fig:microvtmm}, when no unbalanced huge page exists, HMM-v-Huge can maximally reduce address translation overhead, and the hypervisor can accurately identify the hot data of the VM as 4GB and place it in fast memory to achieve optimal performance. However, As the proportion of unbalanced huge pages increases, HMM-v-Huge leads to suboptimal data placement since the hypervisor is unable to distinguish hot pages accurately. For HMM-v-Base, the hypervisor can always detect 4GB of hot data, achieving better performance than with HMM-v-Huge. Furthermore, we uses tiered memory management with FHPM (FHPM-TMM, described in §\ref{sec:impl}) and show that FHPM-TMM achieves the best performance under different ratios of unbalanced huge pages. This is due to FHPM's ability to detect unbalanced huge pages accurately and split them while retaining balanced huge page to reduce address translation overhead. Performance degrades as the ratio of unbalanced huge pages increases, as more hot data is initially placed in the slow memory.

\subsection{Low Savings in Page Sharing}
In the page-sharing scenario, we use the sharing methods as described in §\ref{sec:pageshare} with huge pages enabled. We set up two VMs with 8 cores and 16GB of DRAM each. The two VMs run 8GB random read programs respectively, where 6GB huge pages are frequently accessed and the remaining 2GB huge pages are rarely accessed. For the frequently accessed huge pages, the $PSR$ of the 2GB huge pages is 0 as balanced huge pages and the $PSR$ of the 4GB huge pages is 0.9 as unbalanced huge pages. Thus the frequently accessed memory is actually 2.4GB (the balanced huge pages and the accessed base page regions within the unbalanced huge pages).
The data is first inserted at the base page granularity and then accessed sequentially. The two VMs insert the same data but in a different order. The total memory usage of the two VMs is around 18GB, with each guest OS occupying approximately 1GB. As shown in the Table~\ref{tab:ksm-micro}, huge page sharing does not provide significant memory savings, but it does offer the best performance. Linux KSM can save the most memory, but it splits almost all huge pages, resulting in a performance loss of about 27\%. Ingens divides huge pages into hot and cold, but the hot bloat causes Ingens to incorrectly treat the 6GB of frequently accessed huge pages as all hot memory, which prevents the sharing of the base page regions that are not accessed in unbalanced huge pages. So that it can only save 5.2\% of memory. We also show the result of page sharing system with FHPM (FHPM-Share, described in §\ref{sec:impl}). FHPM-Share can save 29.6\% memory with a performance loss of 5.6\%. Here, we configure FHPM-Share to prioritize memory saving, and the hypervisor can balance memory saving and huge page ratio according to acceptable performance loss, see §\ref{sec:impl}.

\begin{table}[htbp]
\centering
\tabcolsep=0.1cm
  \caption{Memory savings and performance trade-offs of two VMs running the micro benchmark}
  \vspace{-0.3cm}
\begin{tabular}{cccc}
\hline
Policy                           & Memory saving      & Performance & Huge page ratio \\ \hline
\multirow{2}{*}{Huge page share} & \multirow{2}{*}{6MB (\textless{}0.1\%)}                & VM a: 1     & VM a: 100\%     \\
                                 &  & VM b: 1     & VM b: 100\%     \\ \hline
\multirow{2}{*}{Linux KSM}       & \multirow{2}{*}{8568MB (46.4\%)}            & VM a: 0.732 & VM a: 0\%       \\
                                 &            & VM b: 0.721 & VM b: 1\%       \\ \hline
\multirow{2}{*}{Ingens}          & \multirow{2}{*}{952MB (5.2\%)}             & VM a: 0.981 & VM a: 69\%      \\
                                 &             & VM b: 0.982 & VM b: 69\%      \\ \hline
\multirow{2}{*}{Zero page scan}  & \multirow{2}{*}{152MB (0.8\%)}             & VM a: 1     & VM a: 95\%      \\
                                 &             & VM b: 0.995 & VM b: 95\%      \\ \hline
\multirow{2}{*}{FHPM-Share}            & \multirow{2}{*}{5448MB (29.6\%)}            & VM a: 0.944 & VM a: 21\%      \\
&            & VM b: 0.942 & VM b: 21\%      \\ \hline
\end{tabular}
\label{tab:ksm-micro}
\end{table}

\subsection{Current Solutions}
Several studies have noticed the issues of {\it hot bloat}, yet a mature solution remains elusive. Zhu from Facebook proposes the THP Shrinker~\cite{thpshrinker} as a potential solution, which leverages zero scanning to detect untouched parts of unbalanced huge pages, assuming they are initialized to zero. 
Neha Agarwal and Thomas F. Wenisch propose Thermostat~\cite{thermostat} which tackles the issue by splitting huge pages for monitoring purposes, sampling 5\% at a time and collapsing them after monitoring.
Bergman et al. propose HotBox~\cite{ismm22hotbox} which points out that the use of huge pages can lead to improper decisions in a tiered memory management system, but suggests using a base page system instead, which sacrifices the benefits of huge pages in address translation.
Raybuck et al. propose HeMem~\cite{raybuck2021hemem} which uses PEBS~\cite{pebs} to track accessed liner addresses and they declare that these addresses can be counted as page accesses at any granularity. But we find that the accuracy of PEBS is inversely proportional to its overhead, and obtaining the distinguishable access information at base page granularity results in an unacceptable overhead. 
Wang et al. propose Rainbow~\cite{rainbow} which adds address counters to the memory controller to track the frequency of address accesses, but requires complex hardware modifications that are not practical for current cloud services. Thus, there exists a need for a low-overhead, VM-friendly approach that does not require guest OS and hardware modifications.

\section{FHPM design and implementation}

\subsection{Overview}
\begin{figure}[htbp]
\begin{center}
\includegraphics[width=0.48\textwidth]{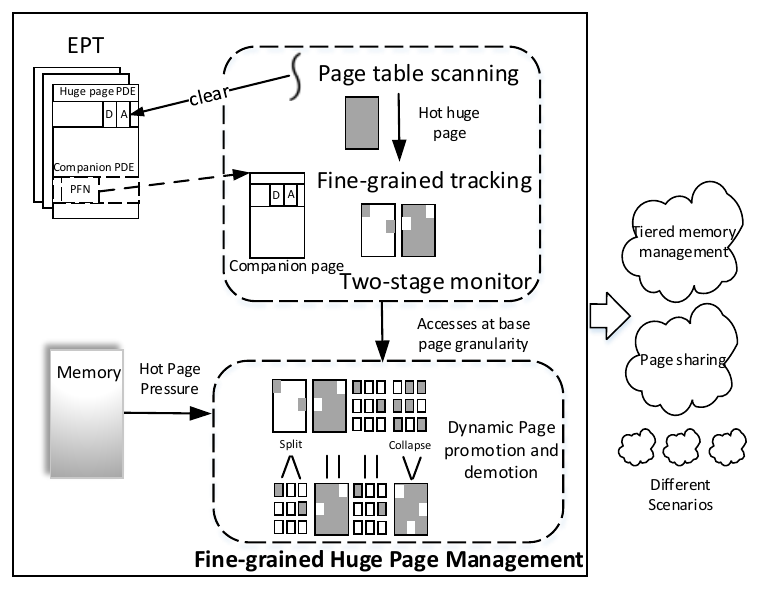}
\caption{The architecture of FHPM. }
\label{fig:all}
\end{center}
\vspace{-0.5cm}
\end{figure}

In this paper, our goal is to address the issue of hot bloat in VMs to balance address translation overhead and memory utilization better. We propose Fine-grained Huge Page Management (FHPM), the overall architecture is shown in Figure~\ref{fig:all}. FHPM aim to obtain access information at base page granularity and dynamically promote/demote pages with low overhead so that the hypervisor can make the proper decisions in different scenarios. 

We employ two-stage monitoring to capture the fine-grained access patterns of the VM. In the first stage, we periodically clear and record A/D bits in EPT to track page access frequency. Subsequently, all the pages are divided into cold and hot pages. For hot huge pages, we redirect their PDEs to the corresponding companion pages. Upon encountering an EPT page walk, the MMU automatically traverses the companion pages to set the A/D bits at the base page granularity. Once the fine-grained access information of the VM is obtained, FHPM dynamically promotes or demotes pages according to the different hot page pressures of the VM instead of a fixed threshold. Moreover, we optimize the page splitting and collapsing methods in Linux to minimize virtualization overheads. Overall, FHPM can complement other memory management subsystems to make better decisions, such as tiered memory management systems and page share systems as discussed before.

\subsection{Fine-grained Tracking}
\label{sec:fine-track}

\subsubsection{Two-stage Monitoring}
Under virtualization, the hypervisor can periodically clear and record the A/D bits of EPT entries to capture the VM's access behavior. However, this approach has drawbacks as clearing the A/D bits invalidates the TLB of the corresponding page and introduces memory access overhead when setting the bits. The overhead incurred by the page scan method is directly proportional to the number of EPT entries and the frequency of A/D bits clearing. Previous studies have shown that frequent A/D bits clearing in base pages can cause up to 29\%~\cite{ingens} of overhead, which is often mitigated by sampling. However, the overhead of monitoring all huge pages is obviously much lower than monitoring base pages, as there are much fewer huge page table entries than base page table entries. Therefore, the monitoring of small page granularity needs to be restrained to reduce the number of monitored entries. 

We employ a two-stage monitoring strategy, commencing with monitoring all pages for brief periods just like the traditional page scan method, followed by partitioning them into hot and cold pages based on their access frequency. The hot huge pages are then subjected to more fine-grained monitoring for a single extended period to identify which base page regions are accessed. The frequency of accessed base page regions is directly inherited from its belonging huge page instead of multiple periods monitoring. Thus, this avoids multiple monitoring of base page granularity and fine-grained monitoring of cold huge pages, greatly reducing the number of monitored page table entries and therefore the overhead.

\subsubsection{Companion Page Redirection}
To obtain fine-grained access information within a huge page, it is necessary to ensure that each base page region has its own separate A/D bits. We maintain a companion page for each PDE of huge pages, which acts as a EPT page table page that temporarily holds base page mappings to the continuous physical memory regions of the huge page. The lifetime of the companion page is completely controlled by FHPM, and is transparent to the EPT page table page management of hypervisor.

\begin{figure}[htbp]
\begin{center}
\includegraphics[width=0.48\textwidth]{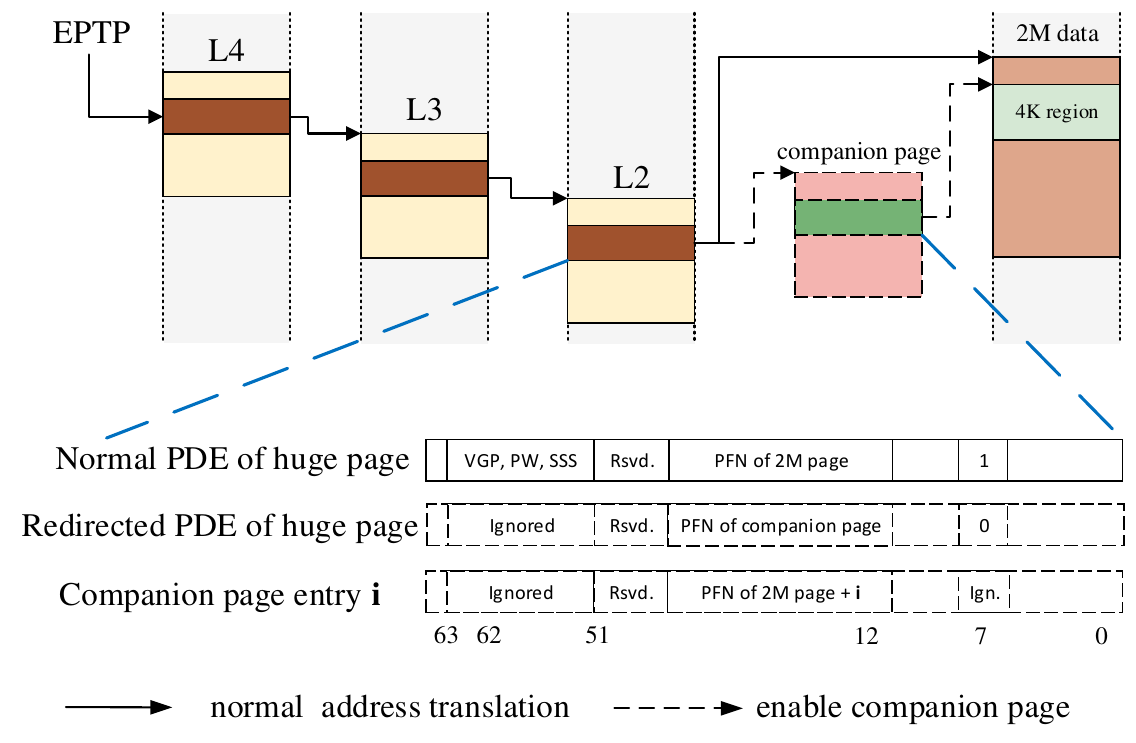}
\caption{Companion page redirection mechanism. Here we take x86 as an example and bit 7 as the page size bit.}
\label{fig:companionpage}
\end{center}
\vspace{-0.5cm}
\end{figure}

Figure~\ref{fig:companionpage} presents the companion page design, which allows for fine-grained monitoring of each base page region within a huge page. Initially, a base page called the companion page is pre-allocated for each PDE of huge pages. Next, the physical address of the companion page associated with the huge page PDE is written to the PFN bits. Additionally, the page size bit in the PDE is needed to switch to allow the MMU to traverse to the companion page based on the PFN in the new PDE during page walks. The companion page contains the corresponding 64-bit page table entries (PTEs) of the base page regions within the huge page. The PFN bits of each entry correspond to the original data PFN, and the permission bits are the same as those in the original PDE. The A/D bits of the corresponding base page region are stored in the PTEs of the companion page. These bits are set by the MMU when the region is accessed or modified. Once monitoring is complete, the companion page is recycled, and the EPT is restored with the original PDE of huge page, and the page size bit is reset, as a graceful fallback. This approach simplifies the implementation by avoiding the need to modify the page structures and also allows for better performance by avoiding page splitting and collapsing overhead.

\subsection{Conflict Resolution}
The redirection of a huge page PDE in the EPT to its companion page creates an inconsistency between the EPT page table and the VM process page table on the host. If the VM process page table is modified by the host OS during monitoring, a VM-exit due to EPT\_VIOLATION is generated, triggering the EPT page fault handler function to walk the EPT and synchronize this change. However, this inconsistency may lead to the abort of the VM. To ensure the safety of the VM, we intercept the monitored pages in the handler when monitoring is activated, restore the original huge page PDE and then synchronize the VM process page table with the EPT.

We utilize a bit for each PDE to indicate if it has been redirected. When monitoring is active, the handler verifies if the PDE associated with the exception has been redirected to its companion page. If such a redirection is detected, the host OS management should take the priority over the hypervisor page monitoring, and thus recycle the companion page of the affected PDE. Additionally, the PFN and page size bits of the PDE are reverted to their initial state before proceeding with the synchronization of the VM process page table. The introduced bits only cause negligible performance and memory overhead of the page fault handler. Excessive conflicts cause erroneous monitoring and increased performance overhead. However, the the mappings of guest physical address to host physical address are relatively steady, and the hypervisor does not actively modify the VM process page table, indicating that inconsistencies between the EPT and the VM process page table rarely result in conflicts. 

\subsection{Dynamic Page Promotion and Demotion}
To balance the address translation overhead and memory utilization, we employ dynamic page promotion and demotion techniques that respond to hot page pressure. 
We introduce a metric, the {\it hot page pressure} ($HP$), to characterize the extent to which the expected memory usage of a VM is exceeded by the amount of frequently accessed pages. Specifically, a positive $HP$ indicates that the capacity of the VM's expected memory is insufficient to accommodate all the hot pages. This may be caused by the hot bloat which is triggered by unbalanced huge pages. A negative $HP$ indicates that there is sufficient space to accommodate more balanced huge pages. This may result from a low huge page ratio in the VM. The ideal value of $HP$ is 0, indicating that all hot pages fit within the VM's memory without any hot bloat, and the address translation advantage of huge pages is maximally exploited. The calculation of $HP$ is as follows:
\[
HP=
\begin{cases}
s_{hot} - s_{tot} \times f_{use} & for\ initialization\\
HP - PSR_{i} \times S_{huge} & for\ demoting\ huge\ page\ i\\
HP + PSR_{i} \times S_{huge} & for\ promoting\ huge\ page\ region\ i
\end{cases}
\]

When the two-stage monitoring is finished, $HP$ is initialized as $s_{hot} - s_{tot} \times f_{use}$, where $s_{hot}$ represents the size of hot pages (obtained in the first stage of the two-stage monitoring), $s_{tot}$ represents the total memory of the VM, and $f_{use}$ denotes the expected memory usage ratio by the hypervisor. The appropriate value of $f_{use}$ depends on the specific scenario and is discussed in detail in §~\ref{sec:impl}. The $HP$ adjusts along with page promotion and demotion according to the untouched memory size within a huge page region to achieve the ideal value of 0. 
Specifically, page demotion can expose the cold base page regions within a hot huge page. As those cold pages are no longer treated as hot, the $HP$ should be decreased to $HP - PSR_{i} \times S_{huge}$. Where the $PSR_{i}$ denotes the PSR of the huge page $i$ being demoted and the $S_{huge}$ represents the size of a huge page. In contrast, the page promotion may cause the cold base pages to collapse into a hot huge page, and the $HP$ should be increased accordingly to $HP + PSR_{i} \times S_{huge}$.

When $HP > 0$, a series of huge pages should be selected to demoting as unbalanced huge pages. We prioritize demoting the pages based on the Page Skew Ratio (PSR) defined in §\ref{sec:hotpage}, starting from larger to smaller. This is because a high PSR implies fewer visited base page regions and, thus, lesser speedup in address translation.
Notably, a high $HP$ may simply be caused by a large mount of hot memory, rather than by unbalanced huge page, so it is necessary to set a lower bound for the PSR of unbalanced huge page. It indicates that huge pages with a PSR lower than the lower bound are always treated as balanced huge pages, as they already contain adequate accessed base pages, and prioritizing address translation overhead in this scenario is essential.
When $HP < 0$, a series of large page regions should be selected as balanced huge page regions, and base pages among them should be promoted. we prioritize promoting pages based on the PSR from small to large, giving priority to the huge page region containing more accessed base pages.

\subsection{VM-friendly Page Splitting and Collapsing}
After deciding which pages need to be promoted or demoted, a series of operations including page splitting and collapsing are required. In the Linux kernel, these operations can be achieved using provided interfaces such as $split\_huge\_page()$ and $collapse\_huge\_page()$. Those interfaces require unmap and remap the page table mapping of the corresponding pages. However, they only operate on the VM process page table, and the mappings in the EPT is invalidated. Remap does not fill the EPT entry until the VM accesses the memory and generates VM-exits to establish EPT mappings, which results in additional performance overhead in virtualization.

We propose a VM-friendly page splitting and collapsing mechanism. To split a huge page, we employ an active EPT refill of base pages after the VM process page table remap. On the other hand, collapsing a huge page only requires an EPT refill on the PMD of the huge page. Our proposed mechanism reduces the performance overhead caused by invalidating EPT entries during page split and collapse operations in virtualization.

\subsection{Implementation}
FHPM is a software solution without hardware and guest OS modifications. FHPM adopts QEMU/KVM~\cite{kvm,qemu} as the virtualization platform. We prototype FHPM in the Linux kernel v5.4.142. We implement FHPM in the KVM module and employ huge pages via Linux's Transparent Huge Page (THP) mechanism, and we refer to a 2MB page as a ``huge page" and a 4KB page as a ``base page".

{\bf Fine-grained Tracking: }We first need to monitor access to all pages using traditional page table scanning, and then redirect the PDEs of hot huge pages to companion pages. We use {\it rmap} in KVM to quickly locate all the PDEs and PTEs of the EPT without having to traverse the entire page table. At the beginning of the companion page redirection, the corresponding {\it mmu\_lock} of the VM is locked to avoid conflicts with EPT access. Then, the {\it PS} bit (bit 7) in the PDE is toggled from 1 to 0 in the  Intel EPT. The A/D bits of the companion page PTE are bit 8 and bit 9, just like the normal PTE in the Intel EPT. After the companion
page redirection is complete, the {\it mmu\_lock} is unlocked and the TLB is flushed.

{\bf Conflict Resolution: }The VM process page table is actually the QEMU process page table managed by the host OS. The changes in QEMU process page table are synchronized with EPT in {\it tdp\_page\_fault()}. So we just need to check whether the corresponding bit of the abnormal PDE is set in the function entrance of {\it tdp\_page\_fault()}. The set PDEs need to fall back to its initial value and unset the bit before returning {\it tdp\_page\_fault()} to continue execution.

{\bf Dynamic Page Promotion and Demotion: } 
Unbalanced huge pages are dynamically selected to be demoted from their PSRs large to small according to the hot page pressure. But there is still a lower bound on the selection of unbalanced huge page. The lower bound of the unbalanced huge page PSR is set to 0.5, which means that a huge page with more than half of the base page regions accessed are always considered as balanced huge page and never to be demoted.

{\bf VM-friendly Page Splitting and Collapsing: } We change the implementation of $split\_huge\_page()$ and $collapse\_huge\_page()$ in Linux.
Specifically, $split\_huge\_page()$ need to (1) check privilege bits, (2) unmap the original huge page PMD, (3) modify the relevant tag bits in struct page and update base pages map count, (4) get a page table page and point the PMD to the page table page, (5) remap the 512 base pages PTEs. All the operates are on the QEMU process page table, we then retrieving a EPT page table page from a vCPU's $mmu\_page\_cache$, removing the original EPT PDE of huge page, and populating the EPT page table page with 512 PTEs.
$collapse\_huge\_page()$ need to (1) allocate a new physical huge page, (2) flush TLB in this 2M virtual address range, (3) unmap base pages and copy them to new huge page, (4) make new huge page PMD and remap it in process page table. Then we reclaim the EPT page table page and modify the page table PDE into a huge page PDE in EPT.

\section{Case Studies}
\label{sec:impl}
Hot bloat can result in incorrect memory usage calculations by page tracking methods, which in turn can lead to sub-optimal decisions in various memory subsystems such as misplaced data in the tiered memory management system and unshared base page regions in the page sharing system. To address this issue, FHPM monitors access to base page regions within huge pages and dynamically promotes or demotes pages based on memory pressure. We present two cases where FHPM makes more informed decisions in memory management subsystems, thereby improving system performance.

{\bf Case study 1:} In the tiered memory system, we implement a tiered memory management for VM with FHPM (FHPM-TMM). We adopt a classical approach for tiered memory management that involves page monitoring, page classification, and page migration. FHPM-TMM is capable of tracking both the base and huge pages at the base page granularity by two-stage monitoring. For huge pages, FHPM-TMM demotes the unbalanced huge page to prevent the cold base page region from occupying the fast memory and promote frequently accessed continuous base pages to a balanced huge page to reduce address translation overhead. After page splitting and collapsing, the balanced huge page and the base pages with high frequency are classified as hot pages and the rest are cold pages. Finally, FHPM-TMM migrate hot pages to the fast memory and cold pages to slow memory. In this case, $ f _ {use} = \frac {s_{fast}}{s_{tot}} $ where $s_{fast}$ is the fast memory size and $s_{tot}$ is the total memory size of the VM. We assume that the total memory is always enough in this scenario due to high capacity of the slow memory. The hypervisor anticipates that the VM can utilize the fast memory to its full potential.

{\bf Case study 2:} In the page sharing system, we implement a page sharing system with FHPM (FHPM-Share). After two-stage monitoring, FHPM-Share divides all huge pages into hot and cold, and hot huge pages are further tagged into unbalanced huge page and balanced huge page. The page splitting and collapsing are delayed because the content check is needed to guarantee memory sharing. FHPM-Share then builds stable and unstable trees ( two red-black trees ) to check same content by scanning all base page regions of VMs, just like KSM. Only cold pages and unbalanced huge pages with shared candidates are actually splitted into base pages. At the same time, only all base pages are not shared within a huge page region can be actually collapsed to a huge page. In this case, the hypervisor typically wants the VM to keep the memory usage below a certain waterline through page sharing to prevent costly swapping or ballooning. FHPM controls this waterline through $f_{use}$, which is usually set to 0.85. At the same time, we also try to prioritize memory savings, that is, setting $f_{use}$ to 0.5, FHPM will split pages more aggressively and pursue more page sharing.
\section{Evaluation}
\label{sec:evaluation}

\subsection{Experimental Setup}
\label{sec:ex-setup}
We evaluate FHPM on a dual-socket Intel Cascade Lake-SP server with 24 cores/48 threads per socket at 2.2GHz. The VMs are configured with 8-cores and we bind all VM CPUs to a single socket. Both the host OS and guest OS run Ubuntu 18.04 with Linux kernel 5.4.142. We use QEMU 3.1.0 as the hypervisor. We build a real DRAM+NVM system for the tiered memory management. We use QEMU to allocate memory from local DRAM as fast memory and  Intel Optane DC PMem\footnote{Although Intel has killed the Optane memory business, PMem remains the only NVM technology suitable for building a real-tiered memory system.} as slow memory, and applications in VMs preferentially allocate fast memory. For the page-sharing system, we set all VMs to exclusively use DRAM only.

\begin{table}[htbp]
\centering
\tabcolsep=0.1cm
  \caption{Benchmark Configurations.}
  \vspace{-0.3cm}
\begin{tabular}{cc}
\hline
Benchmark     & Information                                                                                                                                 \\ \hline
SPEC CPU 2006 & \begin{tabular}[c]{@{}c@{}}429.mcf with default configuration\\  and poor locality\end{tabular}                                             \\
SPEC CPU 2017 & \begin{tabular}[c]{@{}c@{}}657.xz with default configuration\\  and 16GB footprint\end{tabular}                                             \\
Graph 500     & \begin{tabular}[c]{@{}c@{}}Graph 500 v3.0.0 running SSSP and BFS\\  algorithm on a graph with $2^{25}$ points\end{tabular}                  \\
GAPBS         & \begin{tabular}[c]{@{}c@{}}GAPBS running BC and PR algorithm\\  on a Kronecker graph with $2^{25}$ points\end{tabular}                      \\
Redis         & \begin{tabular}[c]{@{}c@{}}Redis v6.0.8 running a YCSB workload \\ with 20GB data, operations follow \\ hotspot distribution\end{tabular}   \\
MongoDB       & \begin{tabular}[c]{@{}c@{}}MongoDB v6.0.5 running a YCSB workload\\  with 20GB data, operations follow\\  hotspot distribution\end{tabular} \\ \hline
\end{tabular}
\label{tab:bench}
\end{table}

To comprehensively evaluate the performance of FHPM, we chose a diverse set of benchmarks with varying access patterns. Specifically, the \texttt{SPEC CPU} test suite measures the compute-intensive portion of real-world applications, with a focus on minimizing I/O and evaluating the performance of the CPU, memory, and compiler~\cite{spec06,spec17}. The \texttt{429.mcf} from \texttt{SPEC 2006}~\cite{spec06} is a classic memory-intensive program with poor locality. For \texttt{SPEC 2017}, we selected \texttt{657.xz} with the larger working set, which is an in-memory compression program with a working set close to 16GB. The \texttt{Graph500}~\cite{graph500} evaluates the ability of computer systems to process large graphics data sets when executing different graph traversal algorithms. We configure \texttt{Graph500} to perform 8 times \texttt{BFS} and \texttt{SSSP} on a graph with $2^{25}$ points (about 20GB). \texttt{GAPBS}~\cite{gap}, a graph computing benchmark suite, is widely used to evaluate the performance of various computer systems and platforms. We run the Page Rank (PR) and Betweenness Centrality (BC) algorithms on a Kronecker graph containing $2^{25}$ points, which takes up to 20GB of memory. In addition, \texttt{Redis}~\cite{redis} and \texttt{MongoDB}~\cite{mongodb} are selected as database benchmarks, and they are tested by using a YCSB~\cite{ycsb} workload, configured with 4KB value size, 20GB data in total, the ratio of READ to UPDATE operations is 1:1, and the operations follows a hotspot distribution, where 80\% of operations occurr on 20\% of data. All benchmarks are summarized in Table~\ref{tab:bench}.

\subsection{Fine-grained Tracking Efficiency}
\label{sub:ex-pagemonitor}
In this section, we present an evaluation of the page monitoring mechanism, focusing on its overhead and accuracy. To this end, we compare FHPM's fine-grained tracking approach against several other methods, including split scan monitoring, sampling scan monitoring, zero scan monitoring, and PEBS. Split scan monitoring involves calling $split\_huge\_page()$ on all huge pages, performing periodic A/D bits clears on the split base pages, and then calling $collapse\_huge\_page()$ to revert to the original huge page state. Sampling scan monitoring is similar to Thermostat, with 5\% of the huge pages selected for splitting. Zero scan monitoring is the same as THP shrinker, which scans the VM data pages to find zero base page regions. It should be noted that PEBS is not suitable for virtualization, as non-root mode cannot write data to the PEBS buffer in the host space. Therefore, we test PEBS directly in the native environment. This is indeed biased towards PEBS as it eliminates the VM-exits caused by the PEBS buffer full that PEBS virtualization may cause.

\subsubsection{Page Monitoring Overhead}

\begin{figure}[htbp]
\begin{center}
\includegraphics[width=0.48\textwidth]{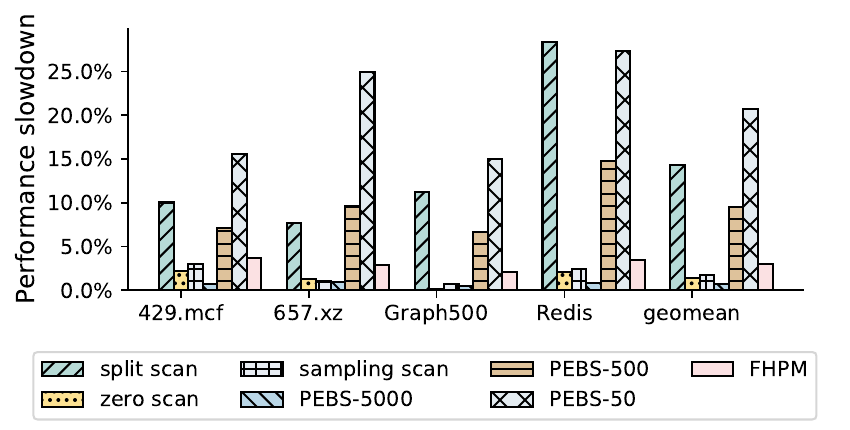}
\caption{Performance slowdown caused by different monitoring mechanisms. }
\label{fig:monitor_slowdown}
\end{center}
\vspace{-0.5cm}
\end{figure}

To evaluate the overhead caused by different monitoring mechanisms, we conducted a thorough evaluation by executing all benchmarks listed in Table~\ref{tab:bench}. Due to space constraints, we only present a subset of benchmark results in this paper, while the results of other benchmarks exhibit similar trends. For all monitoring mechanisms, we enable monitoring for every 10 out of 20 seconds, and within the 10-second monitoring window, the A/D bits clearing interval for the page table scan mechanisms is set to 1 second. The results are shown in Figure~\ref{fig:monitor_slowdown}. The split scan incurs an average performance overhead of 14.32\%, with up to 28.39\% overhead observed for \texttt{Redis} due to the frequent splitting and collapsing of huge pages. The sampling scan has an average performance penalty of 1.78\% as only 5\% of huge pages are selected for splitting and collapsing on each monitor, but this limited sampling rate decreases monitoring accuracy. The zero scan traverses all base page regions to check for zero pages, with an overhead proportional to the number of zero regions and an average performance loss of 1.45\%. PEBS controls the overhead by adjusting the sampling period, with smaller periods resulting in higher accuracy but also higher overhead. We test PEBS with sampling periods of 50, 500, and 5000. The average performance overheads of PEBS-5000, PEBS-500, and PEBS-50 are 0.74\%, 9.51\%, and 20.70\% respectively. Although PEBS-5000 has a low performance overhead, its accuracy is unacceptable. FHPM uses the fine-grained tracking to monitor the access information at the base page granularity, resulting in an average cost of 3.04\%, much lower than the split scan and PEBS-50.

\begin{figure}[htbp]
\begin{center}
\includegraphics[width=0.48\textwidth]{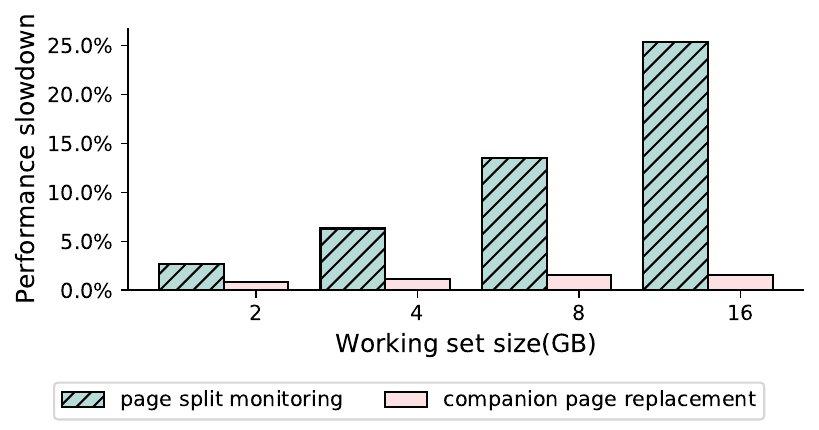}
\caption{Performance slowdown caused by page split monitoring and companion page redirection.}
\label{fig:splitcmp}
\end{center}
\vspace{-0.5cm}
\end{figure}

Periodically clearing A/D bits can cause TLB invalidation and memory access to set A/D bits overhead. So we also tested the performance loss caused by companion page redirection and page split and collapse separately except the overhead caused by clearing A/D bits. Specifically, we used a sequential read/write micro benchmark with a 1:1 read/write ratio and varied the working set size from 2GB to 16GB. We test split scan monitoring and fine-grained tracking of FHPM every 10 out of 20 seconds, but without clearing the A/D bits. As shown in Figure ~\ref{fig:splitcmp}, the performance degradation of page split and collapse is proportional to the size of the working set, with a 16GB working set causing a performance degradation of up to 25.39\%. In contrast, the companion page redirection performs only a lightweight change in EPT, which has a minimal impact on the VM, with a performance overhead of less than 1.5\%.

\subsubsection{Page Monitoring Accuracy}
\begin{figure}[htbp]
\begin{center}
\includegraphics[width=0.48\textwidth]{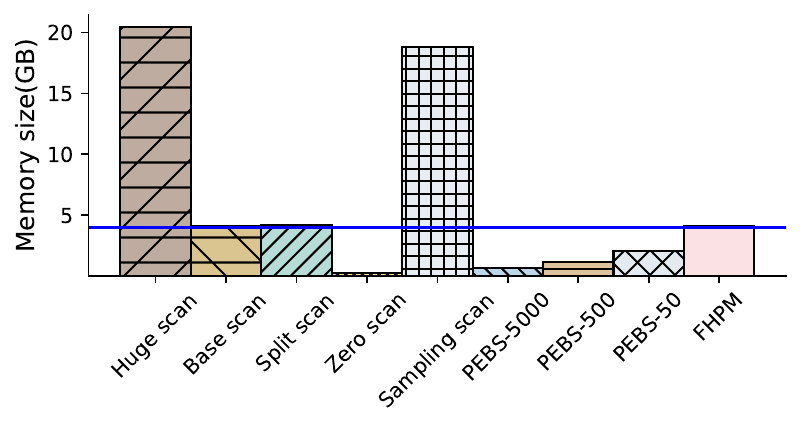}
\caption{Interval monitoring results on \texttt{Redis}. The monitor mechanism is turned on for every 10 out of 20 seconds to get accessed memory size. Base scan is the baseline.}
\label{fig:monitor-acc}
\end{center}
\vspace{-0.5cm}
\end{figure}

To evaluate the accuracy, we use database benchmarks as the workloads because their working set size and access characteristics are predictable, so as to check the accuracy of the monitoring mechanism. We test full monitoring and interval monitoring of the workload respectively. Full monitoring refers to continuous monitoring with A/D bit clearing every 1 second, while interval monitoring involves monitoring for every 10 out of 20 seconds and taking the average value of the memory accessed. We also use base page scan and huge page scan for comparison. Base page scan can track memory accesses at base page granularity, so we use its results as baseline. The results are shown in Table~\ref{tab:fullmonitor} and Figure~\ref{fig:monitor-acc}. Table~\ref{tab:fullmonitor} shows the memory size in different access frequency intervals and the frequency is normalized to 100. Figure~\ref{fig:monitor-acc} shows the interval monitoring results, where the y-axis is the average access memory size monitored by the page monitoring mechanism. 

Due to the hot bloat problem, the full monitoring and interval monitoring results of huge page scan suggest that all memory regions are frequently accessed, which is indeed false. Sampling scan also suffers from hot bloat, as it only samples a small partition of huge pages for splitting and collapsing. In contrast, the full monitoring results obtained by the split scan and the FHPM methods can accurately identify frequently accessed memory regions as the base page scan, and the interval monitoring results indicate that the VM only accesses a small portion of the memory. PEBS obtains the linear address of memory access, which we consider it to be the associated base page access. In the full monitoring, PEBS-5000 and PEBS-500 show that all pages are accessed at a very low frequency, making it impossible to distinguish the cold and hot pages. This limitation arises because the low sampling frequency cannot cover the large number of pages in a typical large working set program. PEBS-50 can detect frequently visited pages closer to the baseline than PEBS-500 and PEBS-5000, but there is still a gap. Finally, the zero scan only counts zero pages, which does not provide an accurate reflection of the VM's exact memory access behavior.

\begin{table}[htbp]
\centering
\tabcolsep=0.1cm
  \caption{Full monitoring Results on Redis.}
  \vspace{-0.3cm}
\begin{tabular}{cccccc}
\hline
\multirow{2}{*}{Monitors} & \multicolumn{5}{c}{Memory size (MB) on frequency intervals}          \\
                          & {[}0,20) & {[}20,40)    & {[}40,60)    & {[}60,80)    & {[}80,100{]} \\ \hline
{\bf Base scan (baseline)}            & {\bf 19171}    & {\bf 184}          & {\bf 79}           & {\bf 11}           & {\bf 1029}         \\
Huge scan            & 330      & 124          & 163          & 1528         & 18186        \\
Split scan                & 19243    & 183          & 82           & 11           & 955          \\
Zero scan                 & 20344    & 3            & 19           & 3            & 1            \\
Sampling scan             & 1144     & 38           & 491          & 975          & 17830        \\
PEBS-5000                 & 20479    & \textless{}1 & \textless{}1 & \textless{}1 & \textless{}1 \\
PEBS-500                  & 19593    & 885          & \textless{}1 & \textless{}1 & \textless{}1 \\
PEBS-50                   & 19185    & 4            & 2            & 2            & 1284         \\
FHPM              & 19271    & 183          & 70           & 10           & 943          \\ \hline
\end{tabular}
\label{tab:fullmonitor}
\end{table}

\subsubsection{Conflicts of Companion Page Redirection}
FHPM uses the PDE pointing to companion pages to replace the huge page PDE in the EPT, thus causing inconsistencies between the EPT and QEMU process page table. Conflicts occur when the host OS manages the QEMU process pages, which lead to VM-exits and monitoring fallbacks. We select a set of benchmarks and use the companion page all the time and count conflicts and the number of $tdp\_page\_fault$ generated during the runtime. The results are shown in table~\ref{tab:conflicts}, which shows that the VM secondary mappings are very stable and the host OS does not usually actively modify or free the VM process page table, so only a few thousand $tdp\_page\_faults$ are generated, and the conflicts caused by the companion page redirection are negligible.

\begin{table}[htbp]
\centering
\tabcolsep=0.1cm
  \caption{Conflicts caused by companion pages redirection.}
  \vspace{-0.3cm}
\begin{tabular}{ccc}
\hline
benchmarks & tdp\_page\_fault & conflicts \\ \hline
429.mcf    & 617              & 4         \\
657.xz     & 5047             & 12        \\
GAPBS-pr    & 590              & 2         \\
GAPBS-bc    & 366              & 3         \\
Graph500   & 1993             & 4         \\
Redis      & 817              & 4         \\
MongoDB    & 2228             & 2         \\ \hline
\end{tabular}
\label{tab:conflicts}
\end{table}

\subsection{Page Promotion and Demotion Efficiency}
\label{sub:ex-pagepromtion}
\begin{figure}[htbp]
\begin{center}
\includegraphics[width=0.48\textwidth]{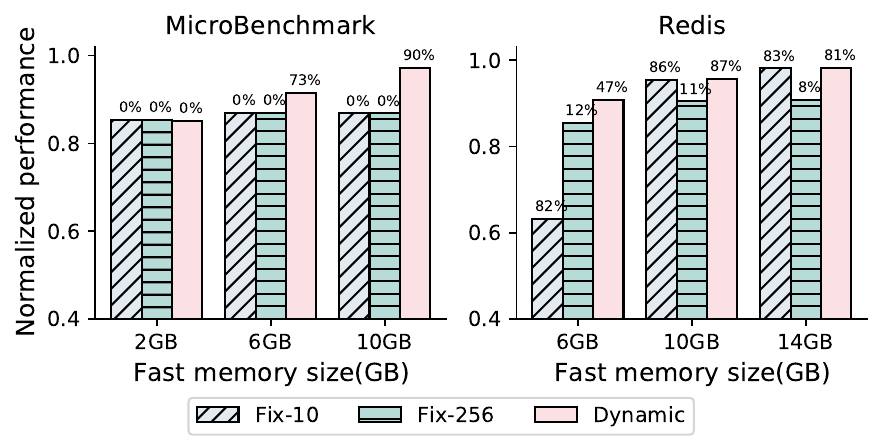}
\caption{The performance of Microbenchmark and \texttt{Redis} in tiered memory system with different fast memory sizes. Percentages on the bars indicate the proportion of huge pages in fast memory.}
\label{fig:dynamic}
\end{center}
\vspace{-0.5cm}
\end{figure}

The state-of-the-art page promotion and demotion policies such as Hawkeye and Ingens use a fixed threshold, whereby a huge page is promoted only after the base page regions used in it reach a fixed amount. In contrast, we propose a dynamic page promotion and demotion policy that takes into account the real-time hot page pressure of the VM, and selects the least overhead-inducing huge page for on-demand promotion or demotion. To evaluate the efficacy of our approach, we test it in a tiered memory management system, where hot page pressure is controlled by the size of fast memory. We vary the size of fast memory and keep the slow memory sufficient. In order to show the limitation of fixed threshold more intuitively, we use an extra microbenchmark to control the access of base page regions in a huge page. The microbenchmark is a 16GB random 1:1 read/write program where each huge page is accessed only 10 base page regions. We also show the results of \texttt{Redis}. For comparison purposes, we utilize fixed thresholds of 10 and 256. Specifically, if the number of base page regions accessed by a huge page region falls below or equal to the threshold, the huge page region is demoted; otherwise, it is promoted. As shown in Figure~\ref{fig:dynamic}, the x-axis represents the size of the fast memory, the bar charts represent the performance of the program relative to running with huge pages enabled with sufficient fast memory, and the percentage of huge pages in the fast memory is shown above the bar.

For the microbenchmark, since each huge page is only accessed 10 base page regions, all huge pages are demoted under 10 and 256 thresholds. But the dynamic policy retains some of the huge pages when the hot page pressure is low, achieving higher performance. For \texttt{Redis}, an aggressive split threshold of 256 solves the hot bloat and achieves a performance of 0.855 when fast memory is 6GB, compared to the performance of 0.633 for the 10 threshold suffering hot bloat. As the fast memory increases to 10GB and 14GB, a threshold of 10 for retaining more huge pages reduces address translation overhead to achieve the performance of 0.955 and 0.981, compared to the performance of 0.904 and 0.908 for the 256 threshold. The dynamic strategy consistently achieves the best trade-off between address translation overhead and memory utilisation, with performance of 0.908, 0.967, 0.982 for 6GB, 10GB, 14GB fast memory respectively. The fixed threshold approach cannot achieve the best performance under all fast memory sizes, whereas our dynamic approach, which accounts for real-time hot page pressure, outperforms it.

\subsection{VM-friendly Page splitting and collapsing Efficiency}
\begin{figure}[htbp]
\begin{center}
\includegraphics[width=0.48\textwidth]{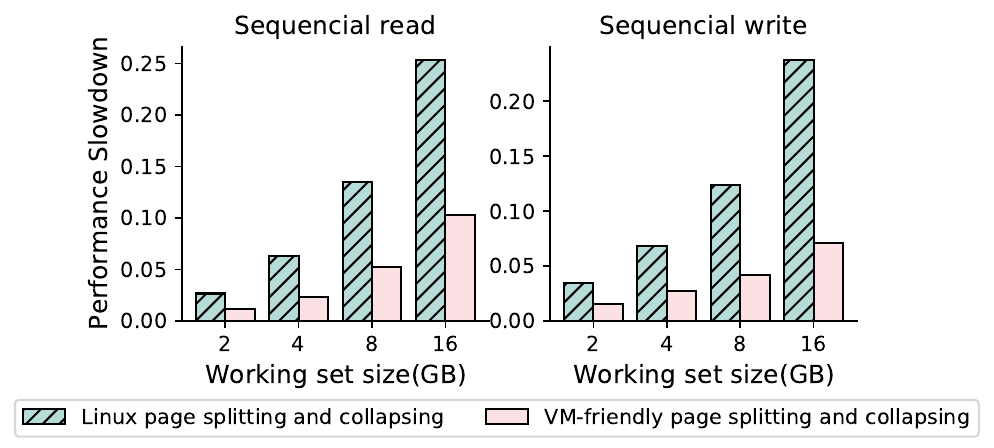}
\caption{The performance slowdown cause by Linux and VM-friendly page splitting and collapsing. }
\label{fig:refill}
\end{center}
\vspace{-0.5cm}
\end{figure}

This section evaluates the VM-friendly page splitting and collapsing mechanism which proactively refilling EPT entries to reduce the number of VM-exits caused by the Linux interfaces. We use a sequential read microbenchmark with varying working set sizes to show it benefits. We split and collapse all pages of the workload every 20s. As shown in figure~\ref{fig:refill}, Linux interfaces causes a slowdown of up to 25.39\% for the 16GB program. With refilling, it is reduced to 10.26\%. In particular, the performance slowdown is still proportional to the working set size, as increasing the working set results in more page table entries and page management data structures to modify. Table~\ref{fig:refill-vmexit} shows that the number of VM-exits caused by VM-friendly page splitting and collapsing, which is much lower than Linux interfaces.

\begin{table}[htbp]
\centering
\tabcolsep=0.1cm
  \caption{VM-exits caused by page splitting and collapsing}
  \vspace{-0.3cm}
\begin{tabular}{ccccc}
\hline
\multirow{2}{*}{WSS(GB)} & \multicolumn{2}{c}{linux interfaces} & \multicolumn{2}{c}{VM-friendly interfaces} \\
                         &  Read              & Write            & Read                & Write                \\ \hline
2                        & 545715            & 547324           & 1197                & 825                  \\
4                        & 1093350           & 1083551          & 1215                & 943                  \\
8                        & 2185320           & 2163738          & 966                 & 977                  \\
16                       & 4355483           & 4285644          & 915                 & 915                  \\ \hline
\end{tabular}
\label{fig:refill-vmexit}
\end{table}

\subsection{Case Study 1: Tiered Memory Management System}
\begin{figure}[htbp]
\begin{center}
\includegraphics[width=0.48\textwidth]{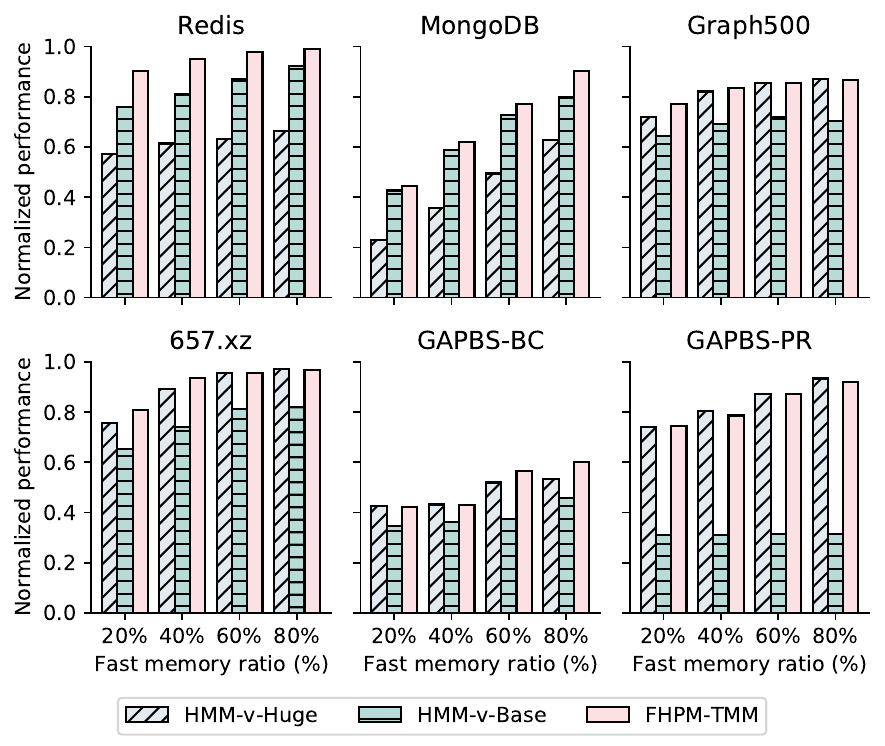}
\caption{
The normalized performance of different fast memory ratio for HMM-v-Huge, HMM-v-Base, and FHPM-TMM. The x-axis represents the ratio of fast memory to the total memory required by the workloads.}
\label{fig:tmmall}
\end{center}
\vspace{-0.5cm}
\end{figure}

For the tiered memory system, we conduct a comparative study of FHPM-TMM with the state-of-the-art solution, HMM-v, of pure huge and base page management (HMM-v-Huge and HMM-v-Base). We show the results of \texttt{Redis}, \texttt{MongoDB}, \texttt{Graph500}, \texttt{657.xz}, and \texttt{GAPBS}. The benchmarks are configured the same as before, and we configure that the VM has enough slow memory but the fast memory size varies. As shown in Figure~\ref{fig:tmmall}, the x-axis represents the ratio of fast memory to the memory required by the benchmarks and the y-axis represents the performance relative to enough fast memory with huge pages enabled. \texttt{Redis} and \texttt{MongoDB} suffer serious hot bloat, so the performance of HMM-v-Huge is lower than that of HMM-v-Base under different fast memory ratios. However, FHPM-TMM outperforms HMM-v-Base because it still retains balanced huge pages to achieve faster address translation. \texttt{Graph500} and \texttt{657.xz} are slightly affected by hot bloat. The performance of HMM-v-Huge is the same as that of FHPM-TMM when the DRAM ratio exceeded 60\% since the fast memory can accommodate all hot base pages and hot huge pages, there are no hot page pressure due to sufficient fast memory. Conversely, when the DRAM ratio is less than 40\%, hot bloats decrease the fast memory utilization. FHPM-TMM dynamically selects unbalanced huge pages to demoting, and achieves the best performance. The performance of HMM-v-Base is lower than that of FHPM-TMM and HMM-v-Huge because the address translation overhead overcomes the fast memory utilization. For \texttt{GAPBS-BC}, FHPM-TMM outperforms HMM-v-Huge when the fast memory is less than 40\%, and is the same as HMM-v-Huge when the fast memory is more than 60\%. This is because the most hot huge pages are all balanced huge pages. When the fast memory is smaller, the balanced huge pages and hot base pages can occupy the entire fast memory. But when increasing the fast memory, the unbalanced huge page have a chance to harm the utilization of fast memory, while FHPM-TMM can demote them and only put the hot base page regions into the fast memory. \texttt{GAPBS-PR} always has no unbalanced huge pages, so the performance of FHPM-TMM is close to that of HMM-v-Huge. The 1\% gap between HMM-v-Huge and FHPM-TMM is due to the monitoring overhead of the companion page. The HMM-v-Base suffers from serious address translation overhead, so its performance is always at a low level.
 
\begin{figure}[htbp]
\begin{center}
\includegraphics[width=0.48\textwidth]{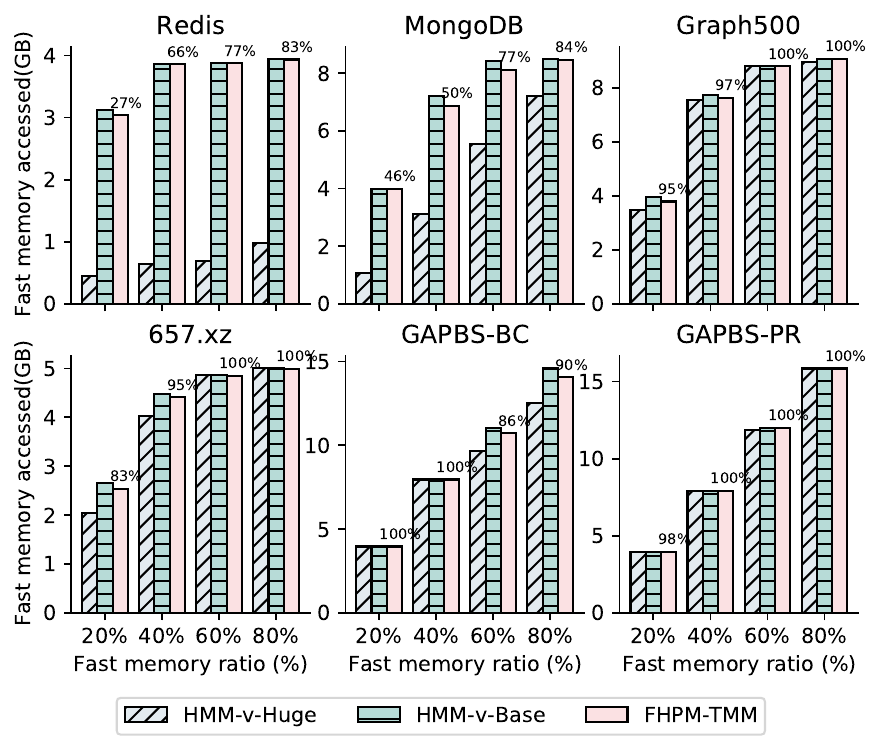}
\caption{The amount of fast memory accessed and the huge page ratio in fast memory. Percentages above the bars indicate the proportion of huge pages in fast memory. Because the huge page ratio of HMM-v-Huge and HMM-v-Base are always 100\% and 0\%, they are not shown in the figure.}
\label{fig:tmmjieshi}
\end{center}
\vspace{-0.5cm}
\end{figure}

We also count the access to fast memory and huge page ratio in fast memory to show the trade-off between address translation overhead and memory utilization, as shown in Figure~\ref{fig:tmmjieshi}. The x-axis represents the ratio of fast memory to the memory required by the benchmarks. The y-axis represents the amount of memory accessed in fast memory, and percentages above the bars indicate the proportion of huge pages in fast memory. Using HMM-v-Huge for \texttt{Redis} and \texttt{MongoDB} leads to low utilization of fast memory, i.e., hot bloat. In contrast, FHPM-TMM can dynamically demote/promote pages based on hot page pressure, thereby achieving the same utilization of fast memory as HMM-v-Base while retaining balanced huge page at the same time. This also explains why the performance of FHPM-TMM is better than HMM-v-Base and HMM-v-Huge. There are fewer unbalanced huge page in \texttt{Graph500} and \texttt{657.xz}. When the fast memory ratio exceeds 60\%, the entire hot region can be accessed in fast memory, while a ratio less than 40\% leads to less usage of HMM-v-Huge compared to HMM-v-Base. There are more hotter balanced huge pages in \texttt{GAPBS-BC}, which can occupy fast memory when the fast memory ratio is less than 40\%, and unbalanced huge page will occupy fast memory when the fast memory ratio is more than 60\%. \texttt{GAPBS-PR} basically has no unbalanced huge pages and FHPM-TMM retains all huge pages.

\subsection{Case Study 2: Page Sharing System}

\begin{table}[htbp]
\centering
\tabcolsep=0.1cm
  \caption{Memory savings and performance trade-off of three VMs running Redis}
  \vspace{-0.3cm}
\begin{tabular}{cccc}
\hline
Policy                           & Memory saving                    & Performance & Huge page ratio \\ \hline
\multirow{3}{*}{Huge page share} & \multirow{3}{*}{321M(0.52\%)}    & VM a: 1     & 100\%           \\
                                 &                                  & VM b: 1     & 100\%           \\
                                 &                                  & VM c: 1     & 100\%           \\ \hline
\multirow{3}{*}{Linux KSM}       & \multirow{3}{*}{28742MB(46.8\%)} & VM a: 0.902 & 4\%             \\
                                 &                                  & VM b: 0.908 & 3\%             \\
                                 &                                  & VM c: 0.897 & 6\%             \\ \hline
\multirow{3}{*}{Ingens}          & \multirow{3}{*}{785MB(1.28\%)}   & VM a: 0.992 & 97\%            \\
                                 &                                  & VM b: 0.998 & 98\%            \\
                                 &                                  & VM c: 0.981 & 98\%            \\ \hline
\multirow{3}{*}{Zero page scan}  & \multirow{3}{*}{384MB(0.56\%)}   & VM a: 0.984 & 96\%            \\
                                 &                                  & VM b: 0.991 & 96\%            \\
                                 &                                  & VM c: 0.983 & 96\%            \\ \hline
\multirow{3}{*}{FHPM-Share-0.5}        & \multirow{3}{*}{25981MB(42.3\%)} & VM a: 0.969 & 25\%            \\
                                 &                                  & VM b: 0.961 & 24\%            \\
                                 &                                  & VM c: 0.962 & 26\%            \\ \hline
\multirow{3}{*}{FHPM-Share-0.85}       & \multirow{3}{*}{9216MB(15\%)}    & VM a: 0.991 & 83\%            \\
                                 &                                  & VM b: 0.985 & 69\%            \\
                                 &                                  & VM c: 0.987 & 71\%            \\ \hline
\end{tabular}
\label{tab:ksm-redis}
\end{table}

For the page sharing system, we conduct experiments on three VMs running \texttt{Redis} with the same configuration as described previously, utilizing a total of 60GB of memory. The VMs insert the same data, but the order of INSERT and UPDATE operations differs. We compare Huge page share, Linux KSM, Ingens, Zero page scan, FHPM-Share-0.85, and FHPM-Share-0.5, as shown in Table~\ref{tab:ksm-redis}. FHPM-Share-0.85 and FHPM-Share-0.5 represent different $f_{use}$ as described in §~\ref{sec:impl}. Huge page share can hardly share memory but has the highest performance because none of the huge pages are demoted. Linux KSM can share all base pages with the identical content, saving 46.8\% of total memory, but all three VMs have a $\sim$10\% performance penalty. Ingens suffers from hot bloat and only saves 1.28\% of memory. Zero page scan can hardly share zero pages, which indicates that the running application rarely initializes the data to zero and then no longer accesses it. FHPM-Share-0.5 shows aggressive page sharing, expecting to achieve 50\% memory usage, and it ends up sharing 42.3\% of the memory, with an average performance loss of 3.6\%. FHPM-Share-0.85 indicates that the hypervisor expects the memory usage below a safe waterline, and ends up with 15\% page sharing, with an average performance overhead of 1.2\%. FHPM-Share can balance between memory saving and performance by adjusting the threshold.

\section{Discussion and Future Work}
{\bf FHPM in native environment:} We implement FHPM in virtualization based on replacing the huge pages' PDEs in the EPT to obtain the base page granularity access information. In the VMs, where a two-dimensional address mapping exists, the memory management of the process only affects the guest page table once the mappings from GFN to PFN are established in the EPT. Therefore, we have opted to redirect the huge pages' PMD in EPT for few conflicts. Companion page redirection can also be applied to process page tables in native environments to monitor base page granularity accesses when enabling huge pages. However, the process's memory management, including page migration and recycling, modifies the process page table directly, causing significant conflicts with the page table redirections. Hence, a supplementary coordination mechanism between FHPM and process memory management is imperative to minimize conflicts and guarantee the precision of monitoring.

{\bf Support for other optimizations:} 
FHPM facilitates the mixed management of huge and base pages through acquisition of access information at base page granularity. In this study, we employ FHPM to extend tiered memory management and page sharing systems. Additionally, FHPM can be implemented in other memory subsystems based on page access information. For instance, in VM hot migration scenarios, it is common to permit memory access during memory copying. As such, tracking dirty page writes during this period is essential. Repeated copies for pages with write are necessary to ensure memory consistency. However, writing to a few base page regions of a huge page necessitates repeated copying of the entire huge page, leading to increased bandwidth occupancy and high overhead. FHPM employs companion page redirection to monitor the write distribution of base page regions within huge pages. For huge pages with unbalanced writes, only the base page regions that have been written need to be recopied, thus expediting page migration.

{\bf Swapping or ballooning under host memory pressure:} 
When the host memory is insufficient, the hypervisor resorts to memory reclamation techniques such as swapping and ballooning to actively reclaim memory from VMs. These techniques involve modifications to the QEMU process page table. The activation of companion pages can potentially cause many conflicts in such scenarios. Nonetheless, our mechanism aims at enhancing the host's memory utilization to obviate the need for costly swapping and ballooning operations. In cases where the host memory remains insufficient and swapping and ballooning are imminent, we 
only monitor the pages not being swapped out.

\section{Conclusion}
This paper presents FHPM, a novel system for managing huge pages in virtualization without guest OS and hardware modifications. FHPM redirects the huge page PDE in EPT to a companion page table page to achieve fine-grained access monitoring and dynamically promotes or demotes pages according to the hot page pressure. We implement FHPM in tiered memory management system and page sharing system, and conduct experiments on real machines. The experiments show that, compared to other recent solutions, FHPM offers better performance in the tiered memory system, and achieves increased memory savings with minimal performance degradation in the page sharing system.

\bibliographystyle{ACM-Reference-Format}
\bibliography{reference}

\end{sloppypar}
\end{document}